\documentclass[a4paper,11pt]{article}

\usepackage{jheppub}
\usepackage{amsfonts}
\usepackage{amsmath}
\usepackage{amssymb}
\usepackage{graphicx}
\usepackage{hyperref}
\usepackage{slashed}
\usepackage{epstopdf}

\title{\boldmath One loop radiative corrections to the translation-invariant noncommutative Yukawa Theory}

\author[a,b]{K. Bouchachia,}
\author[a]{S. Kouadik,}
\author[b]{M. Hachemane}
\author[c]{and M. Schweda}

\affiliation[a]{Facult\'{e} des sciences et de la technologie, Universit\'{e} de M\'{e}d\'{e}a,\\P\^{o}le universitaire, 26000 M\'{e}d\'{e}a, Algeria}
\affiliation[b]{Facult\'{e} de Physique, Universit\'{e} des Sciences et de la Technologie Houari Boumediene,\\B.P. 32, El-Alia, 16111, Bab-Ezzouar Alger, Algeria}
\affiliation[c]{Institute for Theoretical Physics, Vienna University of Technology,\\Wiedner Hauptstrasse 8-10, A-1040 Vienna, Austria}

\emailAdd{bouchachia.karim@univ-medea.dz}
\emailAdd{skouadik@hotmail.com}
\emailAdd{mahachemane@gmail.com}
\emailAdd{mschweda@tph.tuwien.ac.at}

\abstract{We elaborate in this paper a translation-invariant model for fermions in
4-dimensional noncommutative Euclidean space. The construction is done on the
basis of the renormalizable noncommutative translation-invariant $\varphi^{4}$
theory introduced by R. Gurau et al. We combine our model with the scalar
model, in order to study the noncommutative pseudo-scalar Yukawa theory. After
we derive the Feynman rules of the theory, we perform an explicit calculation
of the quantum corrections at one loop level to the propagators and vertices.}

\begin{document} 
\maketitle
\flushbottom

\section{ Introduction}

In the last two decades, a lot of work has been devoted to the study of
noncommutative quantum field theories. The main idea behind these theories is
that at the Planck scale the space-time is no longer commutative, this fact
makes the noncommutative geometry an essential ingredient when probing
spacetime structure at very small distances \cite{doglas}, \cite{zsabo1}. The
original motivation for investigating such theories was the hope of solving
the problem of infinities of quantum field theory and the possible formulation
of consistent quantum gravity \cite{dopl1}, \cite{dopl2}.\ Despite the
collective efforts deployed by physicists none of these goals is yet reached.

In fact, instead of the elimination of ultraviolet infinities, the use of
noncommutative geometry in quantum field theories gives rise to a new set of
problems and makes the short distance behavior of those theories more
ambiguous \cite{filk}-\cite{ramsd}. The conventional theories became non
renormalizable due to the infamous ultraviolet/infrared mixing. Many attempts
were made to overcome this UV/IR mixing, but in general, the problem persists.

However, there are few models in which renormalizability was restored. It\ was
achieved by adding a suitable term to the initial action of the theory. The
procedure was first used by Grosse and Wulkenhaar \cite{gross1}-\cite{gross4}
to solve the UV/IR mixing of the noncommutative Euclidean $\varphi^{4}$
theory. In their model they added a harmonic oscillator term which depends
explicitly on the Moyal space coordinates $\widetilde{x}^{2}\varphi^{2}$,
where $\widetilde{x}_{\mu}=(\theta^{-1})_{\mu\nu}x^{\nu}$. It turns out that
the model is covariant under Langmann-Szabo duality \cite{zsabo2} but also
breaks the translation invariance of the action.

Another approach, using the same method, was proposed by\ R. Gurau et al.
\cite{rivasseau2008}. This model preserves the translation invariance of the
noncommutative $\varphi^{4}$ theory. The term added to the action is in fact a
non local counter-term of the form $\varphi\frac{1}{\theta^{2}\square}\varphi
$, which is written in momentum space as $\frac{1}{\theta^{2}p^{2}}$. This
model is known as the translation-invariant $1/p^{2}$-model. The UV/IR mixing
problem was solved by the elimination of the quadratic IR divergence of
non-planar diagrams. Both of these scalar models were constructed on the Moyal
space and were proven to be renormalizable to all orders in perturbation theory.

The noncommutative fermion theory was also formulated in the case of the
Gross-Neveu model \cite{fabien}. Following the same procedure in Grosse and
Wulkenhaar model, the term added to the action is $\overline{\psi}\gamma^{\mu
}\widetilde{x}_{\mu}\psi$. This model was proven to be renormalizable to all
orders in perturbation theory, but unlike the noncommutative $\varphi^{4}$
models, it still presents a UV/IR mixing even after renormalization. In fact,
the Gross-Neveu model is renormalizable even without adding an extra-term.

Motivated by the renormalizable noncommutative translation-invariant $1/p^{2}%
$-model, and since it has not been extended to fermions, we propose to
construct its fermionic version. It is well known in ordinary quantum field
theory that the scalar propagator is perceived as the square of the Dirac
propagator, indeed we have%

\begin{equation}
\tilde{G}(p^{2})=\frac{1}{p^{2}+m^{2}}=\frac{1}{i%
\slashed{p}%
+m}\times\frac{1}{-i%
\slashed{p}%
+m} \label{cond}%
\end{equation}
this means also that the scalar propagator appears naturally in the expression
of the Dirac propagator%

\begin{equation}
\tilde{D}(p)=\tilde{G}(p^{2})(i%
\slashed{p}%
+m)
\end{equation}
where $\tilde{D}(p)$ is the Dirac propagator and $\tilde{G}(p^{2})$ the scalar
propagator, here expressed in their Euclidean forms.

It seems reasonable to impose the condition (\ref{cond}) in the noncommutative
case if we want to have a consistent theory that involves both scalar and
fermion fields. Thus, our starting point is the construction of a model in
which the modified scalar and fermion propagators are correlated in the same
way as\ in the ordinary quantum field theory. The extra-term in the fermionic
action is chosen accordingly.

The consistency of our model relies on the fulfillment of the condition
(\ref{cond}), but this does not guarantee its renormalizability. This is why
we apply it, in addition to the scalar model, to study the noncommutative
pseudo-scalar Yukawa theory. We recall that the Yukawa interaction between a
pseudo-scalar field $\varphi$ and a Dirac field $\psi$ is represented in the
Euclidean space by the action
\begin{equation}
S[\psi,\bar{\psi},\varphi]=\int dx^{4}g\bar{\psi}\gamma^{5}\psi\varphi
\end{equation}
this interaction is used in the standard model to describe the coupling of
Higgs particle with fermions. The calculation of the quantum corrections at
one loop level enables us to test the consistency of the whole model and its
renormalizability. Further, it reveals more about the behavior of these
modified models and allows us to improve them if necessary.

We note here, that the method used in the renormalizable models gave an
alternative approach to construct noncommutative field theories free of UV/IR
mixing. So, it was natural to extend these models to noncommutative gauge
field theory, hoping to have the same success. But unfortunately this method
failed to solve UV/IR mixing problem, although several promising approaches
were made \cite{tanasa}-\cite{gross gt}. Currently, there is no explicit
procedure to deal with this problem.

The paper is organized as follows: in the next Section we define our model and
derive its Feynman rules. In Section 3 we perform an explicit Feynman graph
calculations at one loop level in order to evaluate the radiative
contributions to the scalar and the fermion propagators and Yukawa and
$\varphi^{4}$ vertices. The Section 4 is devoted to remarks and conclusions.

\section{ The Model}

The realization of noncommutative modified $\varphi^{4}$ models cited above
was achieved by the substitution of the ordinary product between fields by the
Weyl-Moyal star $\star$ product \cite{akfor}%

\begin{equation}
f(x)\star g(x)\equiv e^{\frac{i}{2}\theta^{\mu\nu}\frac{\partial}{\partial
x^{\mu}}\frac{\partial}{\partial y^{\nu}}}f(x)g(y)\mid_{x=y}%
\end{equation}

This approach is considered to be the simplest way to construct a
noncommutative field theory, the coordinates fulfill, the commutation relation%
\begin{equation}
\lbrack x^{\mu},x^{\nu}]_{\star}=x^{\mu}\star x^{\nu}-x^{\nu}\star x^{\mu
}=i\theta^{\mu\nu}%
\end{equation}
where $(\theta^{\mu\nu})$ is the deformation matrix, it is assumed to have a
simple block-diagonal form%
\begin{equation}
(\theta^{\mu\nu})=\theta\left(
\begin{array}
[c]{cccc}%
0 & 1 & 0 & 0\\
-1 & 0 & 0 & 0\\
0 & 0 & 0 & 1\\
0 & 0 & -1 & 0
\end{array}
\right)
\end{equation}
here $\theta$ is the deformation parameter, it is taken to be real and gives
the measure of the strength of noncommutativity. Throughout this paper we use
the Euclidean metric, the Feynman convention $%
\slashed{a}%
=\mathbf{\gamma }^{\mu }a^{\mu }$ and the notation $\tilde{a}^{\mu }=(\theta
^{\mu \nu })a^{\nu }$.

The free scalar action of the translation-invariant $1/p^{2}$-model is
\cite{rivasseau2008}%
\begin{equation}
S_{\star}[\varphi]=\int\limits_{%
\mathbb{R}
^{4}}dx^{4}\frac{1}{2}\left[  \partial^{\mu}\varphi\star\partial^{\mu}%
\varphi+M^{2}\varphi\star\varphi-a^{\prime2}\varphi\star\frac{1}{\theta
^{2}\square}\varphi\right]  \label{scalaraction}%
\end{equation}
in the Euclidean space. In the expression (\ref{scalaraction}), the parameter
$a^{\prime}$ is a real dimensionless constant. The modified scalar propagator
in momentum space is then%
\begin{equation}
\tilde{G}^{\prime}(p^{2},M,a^{^{\prime}})=\frac{1}{p^{2}+M^{2}+\frac
{a^{\prime2}}{\theta^{2}p^{2}}} \label{scalarpropag}%
\end{equation}

In order to recover the modified scalar propagator from the square of the
fermion propagator, as in the commutative theory (\ref{cond}), we propose to
modify the free fermion action in the following way%

\begin{equation}
S_{\star}[\psi,\bar{\psi}]=\int dx^{4}\left[  \bar{\psi}\star%
\slashed{\partial}%
\psi+m\bar{\psi}\star\psi-b^{\prime}\bar{\psi}\star\frac{%
\tilde{\slashed{\partial}}%
}{\theta^{2}\square}\psi\right]  \label{fermionaction}%
\end{equation}
where $b^{\prime}$ is a real dimensionless constant. We have added an
extra-term $b^{\prime}\bar{\psi}\star\frac{%
\tilde{\slashed{\partial}}%
}{\theta^{2}\square}\psi$ to the original fermion action which reads in
momentum space as $\sim \frac{%
\tilde{\slashed{p}}%
}{\theta ^{2}p^{2}}$.

The Yukawa theory in four dimensions Euclidean space includes the $\varphi
^{4}$ self interaction in order to be renormalized, the noncommutative
interaction action is thus%

\begin{equation}
S_{\star}^{int}=\int dx^{4}\left[  \left(  c_{1}\bar{\psi}\gamma^{5}\star
\psi\star\varphi+c_{2}\varphi\star\bar{\psi}\gamma^{5}\star\psi+c_{3}\bar
{\psi}\gamma^{5}\star\varphi\star\psi\right)  +\frac{\lambda}{4!}\left(
\varphi\star\varphi\star\varphi\star\varphi\right)  \right]
\end{equation}
with the use of the trace property of the star product \cite{micu&jab2001}%
\begin{equation}
\int\left(  f\star g\star h\right)  \left(  x\right)  d^{4}x=\int\left(
h\star f\star g\right)  \left(  x\right)  d^{4}x=\int\left(  g\star h\star
f\right)  \left(  x\right)  d^{4}x
\end{equation}
the pseudo scalar Yukawa action reduces to%
\begin{equation}
S_{\star Y}^{int}[\psi,\bar{\psi},\varphi]=\int dx^{4}\left[  g_{1}\bar{\psi
}\gamma^{5}\star\psi\star\varphi+g_{2}\bar{\psi}\gamma^{5}\star\varphi
\star\psi\right]  \label{actyukawa}%
\end{equation}
where $g_{1}=c_{1}+c_{2}$ and $g_{2}=c_{3}.$

The total action of our model reads%
\begin{equation}
S_{\star}^{tot}[\psi,\bar{\psi},\varphi]=S_{\star}[\varphi]+S_{\star}%
[\psi,\bar{\psi}]+S_{\star}^{int}[\psi,\bar{\psi},\varphi]+S_{\star}^{ct}%
[\psi,\bar{\psi},\varphi]
\end{equation}
where $\varphi$ and $\psi$ are the dressed fields and $\psi_{0}$ and $\psi
_{0}$ are the bare fields, we used as usual the substitution%
\begin{equation}
\psi_{0}=\sqrt{Z_{\psi}}\psi\text{ \ \ and \ }\varphi_{0}=\sqrt{Z_{\varphi}%
}\varphi
\end{equation}

The counter-terms action is then%

\begin{align}
S_{\star}^{ct}[\psi,\bar{\psi},\varphi]=  &  \int dx^{4}\frac{1}{2}\left[
\delta_{\varphi}\partial^{\mu}\varphi\star\partial^{\mu}\varphi+\delta
_{M}\varphi\star\varphi-\delta_{a^{\prime2}}\varphi\star\frac{1}{\theta
^{2}\square}\varphi\right]  +\nonumber\\
&  +\int dx^{4}\left[  \delta_{\psi}\bar{\psi}\star%
\slashed{\partial}%
\psi+\delta_{m}\bar{\psi}\star\psi-\delta_{b^{\prime}}\bar{\psi}\star\frac{%
\tilde{\slashed{\partial}}%
}{\theta^{2}\square}\psi\right]  +\nonumber\\
&  +\int dx^{4}\left[  \delta_{g_{1}}\bar{\psi}\gamma^{5}\star\psi\star
\varphi+\delta_{g_{2}}\bar{\psi}\gamma^{5}\star\varphi\star\psi+\frac
{\delta_{\lambda}}{4!}\varphi\star\varphi\star\varphi\star\varphi\right]
\label{actct}%
\end{align}
where the renormalization factors are
\begin{align}
\delta_{\varphi}  &  =Z_{\varphi}-1,\text{ \ \ \ \ \ \ \ \ \ \ \ \ \ \ \ }%
\delta_{\psi}=Z_{\psi}-1\nonumber\\
\delta_{M}  &  =M_{0}^{2}Z_{\varphi}-M^{2},\text{ \ \ \ \ \ \ \ }\delta
_{m}=m_{0}Z_{\psi}-m\nonumber\\
\delta_{\lambda}  &  =\lambda_{0}Z_{\varphi}^{2}-\lambda,\text{
\ \ \ \ \ \ \ \ \ \ }\delta_{g_{i}}=g_{0i}Z_{\psi}Z_{\varphi}^{1/2}%
-g_{i}\nonumber\\
\delta_{a^{\prime2}}  &  =a_{0}^{\prime2}Z_{\varphi}-a^{\prime2},\text{
\ \ \ \ \ \ \ \ \ }\delta_{b^{\prime}}=b_{0}^{\prime}Z_{\psi}-b^{\prime}%
\end{align}

The different actions written above are used next to derive the Feynman rules
for the propagators and vertices.

\subsection{Propagators}

The noncommutative free theory is the same as the commutative one
\cite{micu&jab2001}, the action remains unchanged, and this is due to the
relation%
\begin{equation}
\int\left(  f\star g\right)  \left(  x\right)  d^{4}x=\int\left(  f\cdot
g\right)  \left(  x\right)  d^{4}x \label{starquad}%
\end{equation}

Even when the actions are modified by adding some extra-terms the propagators
are calculated using the same techniques as the ordinary quantum field theory.
The modified scalar propagator in momentum space (\ref{scalarpropag}) is
written as%

\begin{equation}
\tilde{G}^{\prime}(p^{2},M,a)=\frac{1}{p^{2}+M^{2}+\frac{a^{2}}{p^{2}}}%
\end{equation}
where $a=\frac{a^{\prime}}{\theta}$. It is possible to rewrite this
propagator, under a more suitable form \cite{schweda1}, in order to evaluate
the Feynman integrals by the use of the usual mathematical techniques%
\begin{equation}
\frac{1}{p^{2}+M^{2}+\frac{a^{2}}{p^{2}}}=\frac{1}{2}\sum_{\zeta=\pm1}%
\frac{1+\zeta\frac{M^{2}}{2A^{2}}}{p^{2}+\frac{M^{2}}{2}+\zeta A^{2}}\text{ }%
\end{equation}
where $A^{2}=\sqrt{\frac{M^{4}}{4}-a^{2}}$, and if we use Schwinger's
exponential parametrization, with $M>0$ and $a\not =0$, the propagator is then%

\begin{equation}
\tilde{G}^{\prime}(p^{2},M,a)=\frac{1}{2}\sum_{\zeta=\pm1}\left(  1+\zeta
\frac{M^{2}}{2A^{2}}\right)  \int_{0}^{\infty}e^{-\left(  p^{2}+\frac{M^{2}%
}{2}+\zeta A^{2}\right)  \alpha}d\alpha
\end{equation}

The modified fermion propagator is calculated from the action
(\ref{fermionaction}), we obtain%

\begin{equation}
\tilde{D}^{\prime}(p,m,b)=\frac{1}{-i%
\slashed{p}%
+m-ib\frac{%
\tilde{\slashed{p}}%
}{\theta p^{2}}} \label{fermprop}%
\end{equation}
where $b=\frac{b^{\prime}}{\theta}$. This propagator fulfills the condition
(\ref{cond}), or in this case%
\begin{equation}
\frac{1}{p^{2}+m^{2}+\frac{b^{2}}{p^{2}}}=\frac{1}{-i%
\slashed{p}%
+m-ib\frac{%
\tilde{\slashed{p}}%
}{\theta p^{2}}}\times\frac{1}{i%
\slashed{p}%
+m+ib\frac{%
\tilde{\slashed{p}}%
}{\theta p^{2}}}%
\end{equation}
thereafter, the modified scalar propagator is naturally recovered in the
expression of $\tilde{D}^{\prime}$%
\begin{equation}
\tilde{D}^{\prime}(p,m,b)=\tilde{G}^{\prime}(p^{2},m,b)\left(  i%
\slashed{p}%
+m+ib\frac{%
\tilde{\slashed{p}}%
}{\theta p^{2}}\right)  \text{\ }%
\end{equation}
as a consequence, the fermion propagator reproduces the same "damping"
behavior for vanishing momentum as the modified scalar propagator
\cite{schweda1}%
\begin{equation}
\lim_{p\rightarrow0}\tilde{D}^{\prime}(p,m,b)=0
\end{equation}

\subsection{Vertices}

The Feynman rule in momentum space for the $\varphi^{4}$ self interaction
vertex is given by \cite{micu&jab2001}%

\begin{equation}%
\raisebox{-0.2508in}{\includegraphics[
height=0.6391in,
width=0.6443in
]%
{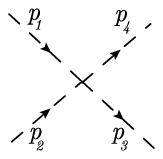}%
}%
=V_{\lambda}(p_{1},p_{2},p_{3},p_{4}) \label{vertexPhi}%
\end{equation}
where $V_{\lambda}(p_{1},p_{2},p_{3},p_{4})=-\frac{\lambda}{3}\left(
\cos\frac{p_{1}\tilde{p}_{2}}{2}\cos\frac{p_{3}\tilde{p}_{4}}{2}+\cos
\frac{p_{1}\tilde{p}_{3}}{2}\cos\frac{p_{2}\tilde{p}_{4}}{2}+\cos\frac
{p_{1}\tilde{p}_{4}}{2}\cos\frac{p_{3}\tilde{p}_{2}}{2}\right)  $, we notice
the factor that appears in the noncommutative case, however in the
commutative limit $\theta\rightarrow0$, it vanishes and we recover the
ordinary $\varphi^{4}$ vertex : $V_{\lambda}\rightarrow-\lambda$.

The Feynman rule for the Yukawa interaction vertex in momentum space is
calculated from the\ Yukawa action (\ref{actyukawa}) using the Fourier
transformation of the fields%

\begin{equation}
\psi(x)=\int\frac{d^{4}p}{\left(  2\pi\right)  ^{4}}\tilde{\psi}%
(p)e^{-ipx},\text{ \ }\bar{\psi}(x)=\int\frac{d^{4}q}{\left(  2\pi\right)
^{4}}\overset{\sim}{\bar{\psi}}(p^{\prime})e^{ip^{\prime}x},\text{ \ }%
\varphi(x)=\int\frac{d^{4}k}{\left(  2\pi\right)  ^{4}}\tilde{\varphi
}(k)e^{-ikx}%
\end{equation}
thus%

\begin{equation}%
\raisebox{-0.2006in}{\includegraphics[
height=0.4877in,
width=0.6849in
]%
{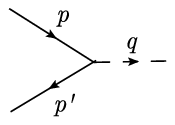}%
}%
=-\gamma^{5}V_{g}(p,p^{\prime})
\end{equation}
where $V_{g}(p,p^{\prime })$ is a phase factor 
\begin{equation}
V_{g}(p,p^{\prime })=\left[ g_{1}e^{+\frac{i}{2}p^{\prime }\tilde{p}%
}+g_{2}e^{-\frac{i}{2}p^{\prime }\tilde{p}}\right] =\sum\limits_{\sigma =\pm
1}g_{\sigma }e^{\sigma \frac{i}{2}p^{\prime }\tilde{p}}\   \label{vertexY}
\end{equation}%
in the last expression we used the notation$\ g_{2}\equiv g_{-1}$. The
Yukawa interaction in our model is represented by two coupling constants $%
g_{1}$ and $g_{2}$, the commutative coupling constant is recovered when $%
\theta \rightarrow 0$: $V_{g}\rightarrow g$ where $g=g_{1}+g_{2}$.

We note, finally, that the modification of the ordinary commutative vertices
is a natural consequence of the introduction of Moyal star product, unlike
the propagators which are modified artificially by adding the extra-terms to
the actions.

\subsection{Counter-terms}

The renormalized Feynman rules can be deduced easily from the counter-terms
action (\ref{actct}), they are written in momentum space as%

\begin{align}%
\raisebox{-0.0502in}{\includegraphics[
height=0.2197in,
width=0.7533in
]%
{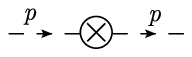}%
}%
&  =-\delta_{\varphi}p^{2}+\delta_{a^{2}}\frac{1}{\theta^{2}p^{2}}-\delta
_{M}\text{, \ \ \ \ \ \ \ \ }%
\raisebox{-0.0502in}{\includegraphics[
height=0.2197in,
width=0.7533in
]%
{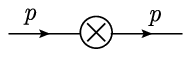}%
}%
=i\delta_{\psi}%
\slashed{p}%
+i\delta_{b}\frac{%
\tilde{\slashed{p}}%
}{\theta^{2}p^{2}}-\delta_{m}\nonumber\\%
\raisebox{-0.2006in}{\includegraphics[
height=0.4877in,
width=0.697in
]%
{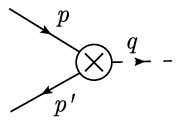}%
}%
&  =-\gamma^{5}\left[  \delta_{g_{1}}e^{+\frac{i}{2}p^{\prime}\tilde{p}%
}+\delta_{g_{2}}e^{-\frac{i}{2}p^{\prime}\tilde{p}}\right]  \text{, \ }%
\raisebox{-0.2508in}{\includegraphics[
height=0.6391in,
width=0.6573in
]%
{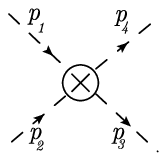}%
}%
=\frac{\delta_{\lambda}}{\lambda}V_{\lambda}(p_{1},p_{2},p_{3},p_{4})
\end{align}

We note from the counter-terms that the constants $a$ and $b$ could receive
corrections in order to eliminate IR divergences of the form $\frac{1}%
{\tilde{p}^{2}}$ and $\frac{%
\tilde{\slashed{p}}%
}{\tilde{p}^{2}}$, respectively. The Feynman rules for propagators and
vertices are now established, they are used in the next section to evaluate
the one loop quantum corrections. The Feynman rules for counter-terms were
also given, they will be used in the renormalization process which will be
discussed in a forthcoming paper.

\section{One loop corrections}

We are going to determine, in this section the relevant corrections for the
1PI two-point functions, for the scalar and fermion field, and the three and
four-point functions at one loop level using dimensional regularization
method. We use the results of the multiscale analysis \cite{rivasseau2008} to
eliminate the subleading logarithmic singularities $\ln\tilde{p}^{2}$ of
non-planar graphs for vanishing momentum, because they represent a mild
divergence \cite{schweda1}. Therefore, we keep in our results only the UV
divergences of the planar integrals and the leading quadratic IR divergences
of the non-planar integrals.

\subsection{Two-point function $\Gamma^{(2)}$}

\subsubsection{Scalar propagator}

The diagrammatic expansion of $\ \Gamma_{\varphi}^{(2)}$ represents the
quantum corrections for the scalar field propagator, at one loop level we have
the tadpole and the fermion loop graphs to evaluate, the first one is
represented by the integral%
\begin{equation}%
{\includegraphics[
height=0.416in,
width=0.8086in
]%
{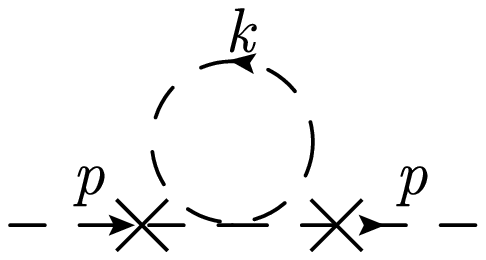}%
}%
=-\frac{\lambda}{6}\int\frac{d^{D}k}{\left(  2\pi\right)  ^{D}}\frac
{1+2\cos^{2}(\frac{p\tilde{k}}{2})}{k^{2}+M^{2}+\frac{a^{2}}{k^{2}}}%
\end{equation}
where the integration is in a $D$-dimension Euclidian space. We evaluate the
divergent part of the planar and non-planar integral using the dimensional
regularization method, the result is
\begin{align}
I_{tadpole} &  =-\frac{\lambda}{6}\frac{1}{\left(  4\pi\right)  ^{2}}%
\sum_{\zeta=\pm1}\left(  1+\zeta\frac{M^{2}}{2A^{2}}\right)  \left(
\frac{M^{2}}{2}+\zeta A^{2}\right)  ^{\frac{D}{2}-1}\Gamma(\frac{2-D}%
{2})-\nonumber\\
&  -\frac{\lambda}{6\left(  4\pi\right)  ^{\frac{D}{2}}}\sum_{\zeta=\pm
1}\left(  1+\zeta\frac{M^{2}}{2A^{2}}\right)  \left(  \frac{\tilde{p}^{2}%
}{4\left(  \frac{M^{2}}{2}+\zeta A^{2}\right)  }\right)  ^{\frac{2-D}{4}%
}K_{1-\frac{D}{2}}\left(  \sqrt{\left(  \frac{M^{2}}{2}+\zeta A^{2}\right)
\tilde{p}^{2}}\right)
\end{align}
where $K_{1-\frac{D}{2}}$ is the modified Bessel function. Thereafter we put
$D=4-\varepsilon$,\ where $\varepsilon\rightarrow0$, this reveals the UV
divergence of the planar part%

\begin{equation}
I_{tadpole}^{P}=\frac{2\lambda M^{2}}{3\left(  4\pi\right)  ^{2}\varepsilon
}+c.c
\end{equation}
the non-planar integral depends on external momentum and it is finite for
$\tilde{p}^{2}\not =0$, however it reveals a leading quadratic IR for
$\tilde{p}^{2}\rightarrow0$
\begin{equation}
I_{tadpole}^{NP}=-\frac{2\lambda}{3\left(  4\pi\right)  ^{2}}\frac{1}%
{\theta^{2}p^{2}}+c.c
\end{equation}

The total divergence of the tadpole integral is then
\begin{equation}
I_{tadpole}=\frac{2}{3}\frac{\lambda M^{2}}{\left(  4\pi\right)
^{2}\varepsilon}-\frac{2}{3}\frac{\lambda}{\left(  4\pi\right)  ^{2}}\frac
{1}{\theta^{2}p^{2}}+O\left(  \lambda^{2}\right)  \label{Stadpol}%
\end{equation}
We note that the UV divergence is different by a numeric factor $\frac{2}{3}$
from the commutative case. This difference is due to the scalar vertex
$V_{\lambda}$ which adds a factor $\frac{1}{3}$ (see reference
\cite{micu&jab2001}) and the extra-term $\frac{a}{k^{2}}$ in the propagator
numerator which adds a factor $2$.

The second contribution to the scalar two-point function comes from the
fermionic loop. After performing a trace over the fermion loop, the integral
representing the second graph reads
\begin{align}%
\raisebox{-0.2309in}{\includegraphics[
height=0.5388in,
width=0.8674in
]%
{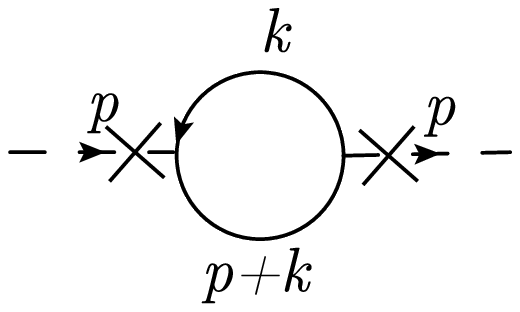}%
}%
&  =-4\int\frac{d^{D}k}{\left(  2\pi\right)  ^{D}}\frac{g_{1}^{2}+g_{2}%
^{2}+2g_{1}g_{2}\cos(p\tilde{k})}{\left(  k^{2}+m^{2}+\frac{b^{2}}{k^{2}%
}\right)  \left(  \left(  p+k\right)  ^{2}+m^{2}+\frac{b^{2}}{\left(
p+k\right)  ^{2}}\right)  }\times\nonumber\\
&  \times\left[  m^{2}+k^{\mu}p^{\mu}+k^{2}-b\frac{k^{\mu}\tilde{p}^{\mu}%
}{\theta k^{2}}+b^{2}\frac{1}{\left(  k+p\right)  ^{2}}+b\frac{k^{\mu}%
\tilde{p}^{\mu}}{\theta\left(  k+p\right)  ^{2}}+b^{2}\frac{k^{\mu}p^{\mu}%
}{k^{2}\left(  k+p\right)  ^{2}}\right]
\end{align}
this can be divided, as usual, into planar and non-planar integrals, following
the same terms order in the last expression, we have%
\begin{equation}
I_{fermion-loop}^{P}=-4\left(  g_{1}^{2}+g_{2}^{2}\right)  \left[  m^{2}%
I_{1}+p^{\mu}I_{2}^{\mu}+I_{3}+b\frac{\tilde{p}^{\mu}}{\theta}I_{4}^{\mu
}+b^{2}I_{5}+b\frac{\tilde{p}^{\mu}}{\theta}I_{6}^{\mu}+b^{2}p^{\mu}I_{7}%
^{\mu}\right]
\end{equation}

and%
\begin{equation}
I_{fermion-loop}^{NP}=-8g_{1}g_{2}\left[  m^{2}J_{1}+p^{\mu}J_{2}^{\mu}%
+J_{3}+b\frac{\tilde{p}^{\mu}}{\theta}J_{4}^{\mu}+b^{2}J_{5}+b\frac{\tilde
{p}^{\mu}}{\theta}J_{6}^{\mu}+b^{2}p^{\mu}J_{7}^{\mu}\right]
\end{equation}
where each integral $I_{i}$ and $J_{i}$ is evaluated separately. For the
planar integrals, we have the following results:

\begin{quote}
-the integrals $I_{1}$, $I_{2}$ and $I_{3}$ presents an UV divergences, their
divergent parts give the contribution $\frac{4\left(  g_{1}^{2}+g_{2}%
^{2}\right)  }{\left(  4\pi\right)  ^{2}\varepsilon}\left(  p^{2}%
+2m^{2}\right)  .$

- the integrals $I_{4}$ and $I_{6}$ are finite, but after integration the
products $\tilde{p}^{\mu }I_{4}^{\mu }$ and $\tilde{p}^{\mu }I_{6}^{\mu }\ $%
are proportional to\ $\tilde{p}^{\mu }p^{\mu }\ $and then vanish.

- the integrals $I_{5}$ and $I_{7}$ are finite for $\theta\not =0$.
\end{quote}

The non-planar integrals are finite for $\tilde{p}^{2}\not =0$, but they could
reveal an IR divergence when $\tilde{p}^{2}\rightarrow0$, we have the
following results:

\begin{quote}
- the integrals $J_{1}$, $J_{2}$ and $J_{3}$ presents an IR divergences, their
divergent parts give the contribution $-\frac{32g_{1}g_{2}}{\left(
4\pi\right)  ^{2}}\frac{1}{\theta^{2}p^{2}}.$

- the integrals $J_{4}$, and $J_{6}$ are finite, but after integration the
products $\tilde{p}^{\mu}J_{4}^{\mu}$ and $\tilde{p}^{\mu}J_{6}^{\mu}\ $are
proportional to\ $\tilde{p}^{\mu}p^{\mu}\ $and then vanish.

- the integrals $J_{5}$, and $J_{7}$ are finite for $\theta\not =0$.
\end{quote}

The fermion contribution to the scalar two-point function reads%

\begin{equation}
I_{fermion-loop}=\frac{4\left(  g_{1}^{2}+g_{2}^{2}\right)  }{\left(
4\pi\right)  ^{2}\varepsilon}\left(  p^{2}+2m^{2}\right)  -\frac{32g_{1}g_{2}%
}{\left(  4\pi\right)  ^{2}}\frac{1}{\theta^{2}p^{2}}+\left(  f_{1}+f_{2}%
p^{2}\right)  +O\left(  g^{3}\right)  \label{SFloop}%
\end{equation}
where $f_{i}$ denote functions that result from the finite integrals, they are
analytic for $\theta\not =0$, this notation is used thereafter. We note here
that the UV divergence in (\ref{SFloop}) is the same as the commutative case
where $g^{2}=g_{1}^{2}+g_{2}^{2}$.

Thus, the total one loop contribution to the scalar field propagator is%

\begin{equation}
\Gamma_{\varphi-1loop}^{(2)}=I_{tadpole}+I_{fermion-loop}%
\end{equation}
$I_{tadpole}\ $and $I_{fermion-loop}$ are given in the expressions
(\ref{Stadpol}) and (\ref{SFloop}), respectively. There is, as expected, a
leading quadratic IR divergence $\sim\frac{1}{\theta^{2}p^{2}}$ resulting from
the non-planar integrals beside the ordinary UV divergence. The additional
term $\left(  f_{1}+f_{2}p^{2}\right)  $ is finite for $\theta\not =0$, it is
a result of the fermionic extra-term, thus it vanishes for $b=0$.

\subsubsection{Fermion propagator}

The quantum corrections for the fermion field propagator at one loop level are
given by the integral%

\begin{align}%
\raisebox{-0.1003in}{\includegraphics[
height=0.4436in,
width=0.7662in
]%
{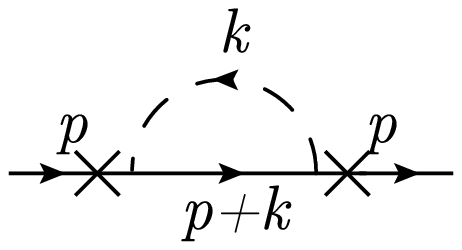}%
}%
&  =-\int\frac{d^{D}k}{\left(  2\pi\right)  ^{D}}\left(  g_{1}^{2}+g_{2}%
^{2}+2g_{1}g_{2}\cos(p\tilde{k})\right)  \times\nonumber\\
&  \times\frac{\left(  i%
\slashed{p}%
-m\right)  +i%
\slashed{k}%
+ib\frac{%
\tilde{\slashed{p}}%
+%
\tilde{\slashed{k}}%
}{\theta\left(  p+k\right)  ^{2}}}{\left(  \left(  p+k\right)  ^{2}%
+m^{2}+\frac{b^{2}}{\left(  p+k\right)  ^{2}}\right)  \left(  k^{2}%
+M^{2}+\frac{a^{2}}{k^{2}}\right)  }%
\end{align}
this can be divided into planar and non-planar integrals, following the same
terms order in the last expression, we have%

\begin{equation}
I^{P}=-\left(  g_{1}^{2}+g_{2}^{2}\right)  \left[  \left(  i%
\slashed{p}%
-m\right)  I_{1}+i\mathbf{\gamma}^{\mu}I_{2}^{\mu}+ib\frac{%
\tilde{\slashed{p}}%
}{\theta}I_{3}+ib\mathbf{\gamma}^{\mu}I_{4}^{\mu}\right]
\end{equation}
and%
\begin{equation}
I^{NP}=-2g_{1}g_{2}\left[  \left(  i%
\slashed{p}%
-m\right)  J_{1}+i\mathbf{\gamma}^{\mu}J_{2}^{\mu}+ib\frac{%
\tilde{\slashed{p}}%
}{\theta}J_{3}+ib\mathbf{\gamma}^{\mu}J_{4}^{\mu}\right]
\end{equation}
We calculate each integral apart and using the same procedure as in the
tadpole integral, we obtain%

\begin{equation}
\Gamma_{\psi-1loop}^{(2)}=-\frac{i\left(  g_{1}^{2}+g_{2}^{2}\right)
}{\left(  4\pi\right)  ^{2}\varepsilon}\left(
\slashed{p}%
+2im\right)  +f_{3}\frac{%
\tilde{\slashed{p}}%
}{\theta}+O\left(  g^{3}\right)  \label{tpff}%
\end{equation}
we note that the UV divergence, in the last relation, is the same as the
commutative theory with $g^{2}=g_{1}^{2}+g_{2}^{2}$. We have also here an
additional term $f_{3}\frac{%
\tilde{\slashed{p}}%
}{\theta}$ which is finite for $\theta\not =0$.

\subsection{Three-point function $\Gamma^{(3)}$}

The one loop quantum corrections to Yukawa vertex are given by only one graph, namely%

\begin{align}%
\raisebox{-0.2914in}{\includegraphics[
height=0.6676in,
width=0.7256in
]%
{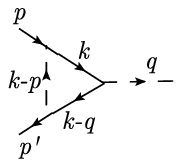}%
}%
&  =-\gamma^{5}\int\frac{d^{D}k}{\left(  2\pi\right)  ^{D}}\frac
{F(k,q,p)N\left(  k,q\right)  }{\left(  k^{2}+m^{2}+\frac{b^{2}}{k^{2}%
}\right)  }\times\nonumber\\
&  \times\frac{1}{\left(  \left(  k-q\right)  ^{2}+m^{2}+\frac{b^{2}}{\left(
k-q\right)  ^{2}}\right)  \left(  \left(  k-p\right)  ^{2}+M^{2}+\frac{a^{2}%
}{\left(  k-p\right)  ^{2}}\right)  } \label{3pnt}%
\end{align}
where $F(k,q,p)$ represents the product of the phase factors of the Yukawa vertices%

\begin{align}
F(k,q,p)  &  =\sum\limits_{\sigma=\pm1}g_{\sigma}e^{\frac{i\sigma}{2}\left[
k\left(  \tilde{q}-\tilde{p}\right)  +q\tilde{p}\right]  }\sum\limits_{\sigma
^{\prime}=\pm1}g_{\sigma^{\prime}}e^{\frac{i\sigma^{\prime}}{2}k\tilde{p}}%
\sum\limits_{\sigma^{\prime\prime}=\pm1}g_{\sigma^{\prime\prime}}%
e^{\frac{i\sigma^{\prime\prime}}{2}k\tilde{p}}\nonumber\\
&  =\sum\limits_{\sigma=\pm1}\left\{  g_{1}g_{2}g_{\sigma}e^{\frac{i\sigma}%
{2}q\tilde{p}}+g_{1}g_{2}g_{\sigma}\left(  e^{i\sigma\left[  k\left(
\tilde{q}-\tilde{p}\right)  +\frac{q\tilde{p}}{2}\right]  }+e^{i\sigma\left(
k\tilde{p}-\frac{q\tilde{p}}{2}\right)  }\right)  +g_{\sigma}^{3}%
e^{i\sigma\left(  k\tilde{q}+\frac{q\tilde{p}}{2}\right)  }\right\}
\end{align}
the first term is independent of $k$, therefore it appears as a factor of the
planar integrals while the other terms enter in the non-planar integrals. The
function $N\left(  k,q\right)  $ represents the product of the fermions
propagators numerators with the $\gamma^{5}$ matrix%
\begin{equation}
N\left(  k,q\right)  =\left[  i%
\slashed{k}%
+m+ib\frac{%
\tilde{\slashed{k}}%
}{\theta k^{2}}\right]  \gamma^{5}\left[  i\left(
\slashed{k}%
-%
\slashed{q}%
\right)  +m+ib\frac{%
\tilde{\slashed{k}}%
-%
\tilde{\slashed{q}}%
}{\theta\left(  k-q\right)  ^{2}}\right]  \gamma^{5}%
\end{equation}
the use of dimensional power counting, reveals that all the terms of this
function contribute in a convergent integrals except the one with $k^{2}$, the
resulting divergence, from the planar and non-planar integrals, is then
logarithmic. Thus, the evaluation of the divergent parts of the integral
(\ref{3pnt}) gives%

\begin{equation}
\Gamma_{1loop}^{(3)}=-\gamma^{5}\left(  \frac{2g_{1}g_{2}}{\left(
4\pi\right)  ^{2}\varepsilon}+f_{4}\right)  \left(  g_{1}e^{\frac{i}{2}%
q\tilde{p}}+g_{2}e^{-\frac{i}{2}q\tilde{p}}\right)  +f_{5}+O(g^{4})
\end{equation}
where $f_{i}$ are analytic functions for $\theta\not =0$ resulting from the
finite integrals. In order to recover results of the UV divergence of the
commutative case we have to make the substitution $g_{1}g_{2}(g_{1}%
+g_{2})=g^{3}.\,$

\subsection{Four-point function $\Gamma^{(4)}$}

In order to evaluate the four-point function at one loop level, we have to
include all the contributions that give $O(\lambda^{2})$ and $O(g^{4})$
corrections to the $\varphi^{4}$ vertex.

\subsubsection{$\Gamma^{(4)}$ with $\varphi^{4}$ coupling}

The scalar one loop contributions to the $\varphi^{4}$ vertex comes from the
following graphs%

\begin{equation}
\Gamma_{\varphi^{4}-1loop}^{(4)}=\frac{1}{2}\left(
\raisebox{-0.3009in}{\includegraphics[
height=0.6849in,
width=0.627in
]%
{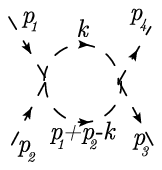}%
}%
+%
\raisebox{-0.3009in}{\includegraphics[
height=0.6374in,
width=0.7109in
]%
{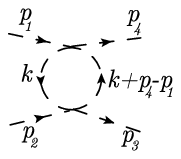}%
}%
+%
\raisebox{-0.3009in}{\includegraphics[
height=0.6374in,
width=0.7109in
]%
{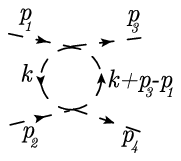}%
}%
\right)
\end{equation}
which are evaluated from these integrals%

\begin{align}
\Gamma_{\varphi^{4}-1loop}^{(4)}  &  =\frac{1}{2}\int\frac{d^{D}k}{\left(
2\pi\right)  ^{D}}F_{1}(k,p_{i})\frac{1}{\left(  \left(  p_{1}+p_{2}-k\right)
^{2}+M^{2}+\frac{a^{2}}{\left(  p_{1}+p_{2}-k\right)  ^{2}}\right)  \left(
k^{2}+M^{2}+\frac{a^{2}}{k^{2}}\right)  }+\nonumber\\
&  +\frac{1}{2}\int\frac{d^{D}k}{\left(  2\pi\right)  ^{D}}F_{2}(k,p_{i}%
)\frac{1}{\left(  \left(  k+p_{4}-p_{1}\right)  ^{2}+M^{2}+\frac{a^{2}%
}{\left(  k+p_{4}-p_{1}\right)  ^{2}}\right)  \left(  k^{2}+M^{2}+\frac{a^{2}%
}{k^{2}}\right)  }+\nonumber\\
&  +\frac{1}{2}\int\frac{d^{D}k}{\left(  2\pi\right)  ^{D}}F_{3}(k,p_{i}%
)\frac{1}{\left(  \left(  k+p_{3}-p_{1}\right)  ^{2}+M^{2}+\frac{a^{2}%
}{\left(  k+p_{3}-p_{1}\right)  ^{2}}\right)  \left(  k^{2}+M^{2}+\frac{a^{2}%
}{k^{2}}\right)  }%
\end{align}
where $F_{i}(k,p_{j})$ is the product of the two $\varphi^{4}$ vertices.

These integrals were evaluated in \cite{schweda1} by introducing a cut-off, we
find equivalent results using the dimensional regularization method. These
diagrams present a logarithmic UV divergence%

\begin{equation}
\Gamma_{\varphi^{4}-1loop}^{(4)}=-\frac{2\lambda}{\left(  4\pi\right)
^{2}\varepsilon}V_{\lambda}(p_{1},p_{2},p_{3},p_{4})+O(\lambda^{3})
\end{equation}
which is different by a numeric factor $\frac{2}{3}$ from the commutative case
where $V_{\lambda}\rightarrow-\lambda$.

\subsubsection{$\Gamma^{(4)}$ with Yukawa coupling}

The contributions to the $\varphi^{4}$ vertex come in this case from the
Yukawa interaction, it represent the fermion corrections to the scalar
$\varphi^{4}$ coupling constant. In order to have an effective contribution to
the $\varphi^{4}$ vertex, from the fermion loop, we need to recover in our
final result the $\varphi^{4}$ extra-factor of $V_{\lambda}$ from the product
of the Yukawa phase factors $V_{g}$. If we consider only the permutations of
external momenta, as in the commutative theory, then we will have only six
diagrams to evaluate, namely%

\begin{equation}
\Gamma_{Y-1loop}^{(4)}=%
\raisebox{-0.3009in}{\includegraphics[
height=0.7403in,
width=0.71in
]%
{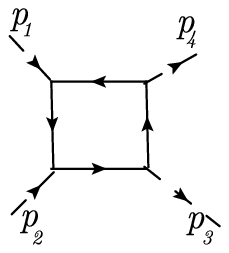}%
}%
+%
\raisebox{-0.3009in}{\includegraphics[
height=0.7403in,
width=0.71in
]%
{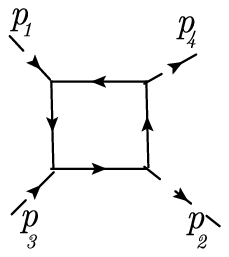}%
}%
+%
\raisebox{-0.3009in}{\includegraphics[
height=0.7403in,
width=0.71in
]%
{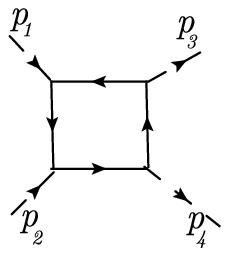}%
}%
+%
\raisebox{-0.3009in}{\includegraphics[
height=0.7403in,
width=0.6944in
]%
{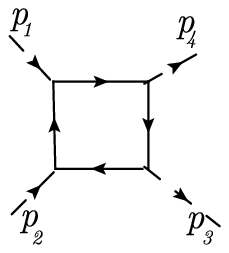}%
}%
+%
\raisebox{-0.3009in}{\includegraphics[
height=0.7403in,
width=0.6944in
]%
{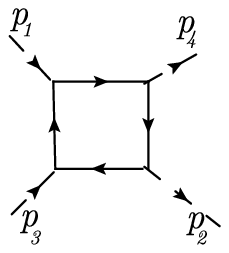}%
}%
+%
\raisebox{-0.3009in}{\includegraphics[
height=0.7403in,
width=0.6944in
]%
{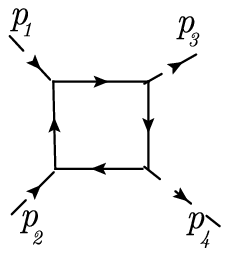}%
}%
\end{equation}

However, in the noncommutative case, there are also the phase factors coming
from the Yukawa vertices which depend explicitly on internal momenta. This
means that when we expand the four-point function at one loop level we will
have more diagrams to evaluate. In fact there only two different permutations
of internal momentum for each one of the last six diagrams, it can be
represented as follow%

\begin{equation}%
\raisebox{-0.3009in}{\includegraphics[
height=0.7403in,
width=0.71in
]%
{phifour3a.eps}%
}%
\rightarrow%
\raisebox{-0.3009in}{\includegraphics[
trim=0.000000in 0.000000in -0.073037in 0.000000in,
height=0.7429in,
width=0.7126in
]%
{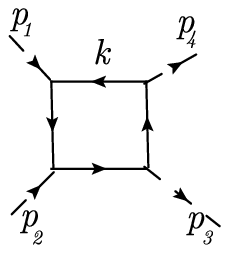}%
}%
+%
\raisebox{-0.3009in}{\includegraphics[
trim=0.000000in 0.000000in -0.072077in 0.000000in,
height=0.7429in,
width=0.7126in
]%
{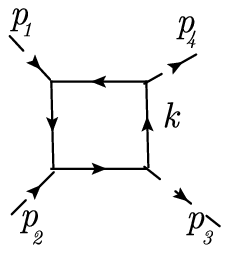}%
}%
\end{equation}

As a result, the diagrammatic expansion of the four-point function is
represented by twelve diagrams, the integral corresponding to each diagram
have the generic form%

\begin{align}
I_{j}^{(4)}  &  =\left(  -1\right)  \int\frac{d^{D}k}{\left(  2\pi\right)
^{D}}\frac{F^{\prime}(k,p_{i})}{\left(  k^{2}+m^{2}+\frac{b^{2}}{k^{2}%
}\right)  \left(  \left(  k+p_{1}\right)  ^{2}+m^{2}+\frac{b^{2}}{\left(
k+p_{1}\right)  ^{2}}\right)  }\times\nonumber\\
&  \times\frac{Tr[N^{\prime}(k,p_{i})]}{\left(  \left(  k+p_{1}+p_{2}\right)
^{2}+m^{2}+\frac{b^{2}}{\left(  k+p_{1}+p_{2}\right)  ^{2}}\right)  \left(
\left(  k+p_{4}\right)  ^{2}+m^{2}+\frac{b^{2}}{\left(  k+p_{4}\right)  ^{2}%
}\right)  } \label{integral4pf}%
\end{align}
where $F^{\prime}(k,p_{i})=\prod\limits_{i=1}^{4}V_{g}^{i}$ is the product of
the phase factors of the four Yukawa vertices it can be written as
\begin{align}
F^{\prime}(k,p_{i})  &  =\sum\limits_{\sigma=\pm1}g_{\sigma}^{4}%
e^{i\frac{\sigma}{2}\left(  p_{m}\tilde{p}_{n}+p_{r}\tilde{p}_{s}\right)
}+g_{1}g_{2}\sum\limits_{\sigma=\pm1}\sum\limits_{\alpha=1}^{4}g_{\sigma}%
^{2}e^{-i\sigma\varepsilon_{\alpha}k\tilde{p}_{\alpha}}e^{\sigma
\varepsilon_{\alpha}\frac{i}{2}\left(  \varepsilon_{\alpha+1}p_{m}\tilde
{p}_{n}+\varepsilon_{\alpha-1}p_{r}\tilde{p}_{s}\right)  }+\nonumber\\
&  +g_{1}g_{2}\sum\limits_{\sigma=\pm1}\sum\limits_{\alpha=2}^{4}g_{1}%
g_{2}e^{\sigma ik\left(  \varepsilon_{\alpha}\tilde{p}_{1}+\tilde{p}_{\alpha
}\right)  }e^{\sigma\frac{i}{2}\left(  p_{m}\tilde{p}_{n}-\varepsilon
_{\alpha-1}p_{r}\tilde{p}_{s}\right)  } \label{phas}%
\end{align}
\ here $\varepsilon _{0,1,2}=1$ and $\varepsilon _{3,4,5}=-1$. The function $%
F^{\prime }(k,p_{i})$ divides the integral (\ref{integral4pf}) into planar
and non-planar parts, the first term in (\ref{phas}) is just a factor of
planar integrals while the others enter in the non-planar integrals. The
indices $m,$ $n,$ $r$ and $s$ take different values from $1$ to $4$, this
gives us twelve different phase factors for $e^{i\frac{\sigma }{2}\left(
p_{m}\tilde{p}_{n}+p_{r}\tilde{p}_{s}\right) }$. These phase factors result
from the permutations of the $p_{i}$, where each one of them correspond to a
different graph.

The trace of the fermionic loop is $Tr[N^{\prime}(k,p_{i})]$, where
\begin{align}
N^{\prime}(k,p_{i})  &  =\gamma^{5}\left(  i%
\slashed{k}%
+m+ib\frac{%
\tilde{\slashed{k}}%
}{\theta k^{2}}\right)  \gamma^{5}\left(  i\left(
\slashed{k}%
+%
\slashed{p}%
_{1}\right)  +m+ib\frac{\left(
\tilde{\slashed{k}}%
+%
\tilde{\slashed{p}}%
_{1}\right)  }{\theta\left(  k+p_{1}\right)  ^{2}}\right)  \times\nonumber\\
&  \times\gamma^{5}\left(  i\left(
\slashed{k}%
+%
\slashed{p}%
_{1}+%
\slashed{p}%
_{2}\right)  +m+ib\frac{\left(
\tilde{\slashed{k}}%
+%
\tilde{\slashed{p}}%
_{1}+%
\tilde{\slashed{p}}%
_{2}\right)  }{\theta\left(  k+p_{1}+p_{2}\right)  ^{2}}\right)
\times\nonumber\\
&  \times\gamma^{5}\left(  i\left(
\slashed{k}%
+%
\slashed{p}%
_{4}\right)  +m+ib\frac{\left(
\tilde{\slashed{k}}%
+%
\tilde{\slashed{p}}%
_{4}\right)  }{\theta\left(  k+p_{4}\right)  ^{2}}\right)
\end{align}

When expanding the product $F^{\prime}(k,p_{i})\times Tr[N^{\prime}(k,p_{i}%
)]$, the integral (\ref{integral4pf}) is divided into thousand of integrals,
fortunately, most of them are finite. The divergent integrals are only those
having $k^{4}$ $\allowbreak$in the numerator, the resulting divergence is then
logarithmic. The summation over all the planar graphs allows us to recover the
$\varphi^{4}$ extra-factor of $V_{\lambda}$
\begin{equation}
\sum\limits_{perms.\text{ }of\text{ }p_{i}}e^{\frac{i}{2}\left(  p_{m}%
\tilde{p}_{n}+p_{r}\tilde{p}_{s}\right)  }=4\left(  \cos\frac{p_{1}\tilde
{p}_{2}}{2}\cos\frac{p_{3}\tilde{p}_{4}}{2}+\cos\frac{p_{1}\tilde{p}_{3}}%
{2}\cos\frac{p_{2}\tilde{p}_{4}}{2}+\cos\frac{p_{1}\tilde{p}_{4}}{2}\cos
\frac{p_{3}\tilde{p}_{2}}{2}\right)
\end{equation}
thus, the fermionic contributions to the scalar coupling constant is%

\begin{equation}
\Gamma_{Y-1loop}^{(4)}=\sum_{j=1}^{12}I_{j}^{(4)}=\left(  \frac{96\left(
g_{1}^{4}+g_{2}^{4}\right)  }{\left(  4\pi\right)  ^{2}\varepsilon}%
+f_{6}\right)  \frac{V_{\lambda}(p_{1},p_{2},p_{3},p_{4})}{\lambda}%
+f_{7}+O(g^{5})
\end{equation}
where $f_{i}$ are analytic functions for $\theta \not=0$ resulting from the
finite integrals. Since twelve graphs are evaluated in the noncommutative
case instead of six, the UV divergence in this case is twice that of the
commutative theory, where $g^{4}=g_{1}^{4}+g_{2}^{4}$.

We note here the importance of recovering the $\varphi ^{4}$ vertex from the
product of Yukawa vertices because it can be seen as a consistency test for
our model.

\section{Conclusions and remarks}

In this work we have constructed a translation-invariant noncommutative
pseudo-scalar Yukawa model and calculated the quantum corrections at one loop
level up to 1PI four-point function. The results obtained will be used
thereafter to discuss the issue of renormalizability for this model and to
adjust it if necessary. However the renormalization process at one loop level
will be discussed in a forthcoming paper.

The analytic functions $f_{i}$ that appear in the some results of quantum
corrections do not affect the renormalizability of our model in the
noncommutative case. Moreover, they can contribute as noncommutative
corrections to the fields, masses and coupling constants. However the
commutative limit could be problematic since $f_{i}\longrightarrow\infty$. In
this case, one has to use the mechanism described in \cite{magnen}, which
relies on the analysis of the UV/IR mixing in Feynman graphs to recover the
commutative theory. The commutative limit for the modified noncommutative
models is not recovered simply by taking $\theta\longrightarrow0$, even with
this mechanism the limit is not smooth.

The presence of the term $\sim\frac{%
\tilde{\slashed{p}}%
}{\theta}$ instead of $\sim\frac{%
\tilde{\slashed{p}}%
}{\tilde{p}^{2}}$ in the fermion two-point function corrections adds a
divergence that cannot be absorbed by any renormalization factor. This
suggests adding, beside the term $\sim\frac{%
\tilde{\slashed{p}}%
}{\tilde{p}^{2}}$, another term of the form $\sim\frac{%
\tilde{\slashed{p}}%
}{\theta}$ to the fermion action. This fact can be explained by the existence
of inner derivative on Moyal space which is different than the one defined on
ordinary space \cite{wallet}. We note that the analytic functions $f_{i}$ and
the term $\sim\frac{%
\tilde{\slashed{p}}%
}{\theta}$ discussed above result from the fermionic extra-term, thus they
vanish for $b=0$. In this case the theory is renormalizable at one loop level
but it doesn't fulfills the consistency condition (\ref{cond}).

Finally, this model can be extended to the gauge field theory as it has been
done with the scalar models, in this case one has to respect BRS symmetry.
However  the loop corrections are harder to evaluate due to the existence of
the extra-terms both in the scalar and fermion actions. The model can be\
also extended to supersymmetry, where this work can be included in the
bosonic part of the theory.

\begin{acknowledgments}
K. Bouchachia acknowledges the financial support of the University of
M\'{e}d\'{e}a for his visits to the Paris-XI University and Vienna University
of Technology. He would like to thank Professors V. Rivasseau and M. Schweda
for their kind invitations and for their help, and also Professor H. Grosse
for his enlightening discussions.
\end{acknowledgments}

\end{document}